\documentclass[a4paper,11pt]{article}

\usepackage{jheppub}
\usepackage{bm}
\usepackage{color}
\usepackage[usenames,dvipsnames,svgnames,table]{xcolor}
\usepackage{amsmath}
\usepackage{amssymb}
\usepackage{graphicx}
\usepackage{slashed}
\usepackage{soul}
\usepackage{multirow}

\def\beq{\begin{equation}}
\def\eeq{\end{equation}}
\def\bea{\begin{eqnarray}}
\def\eea{\end{eqnarray}}
\def\ben{\begin{enumerate}}
\def\een{\end{enumerate}}

\def\lsim{\mathrel{\raise.3ex\hbox{$<$\kern-.75em\lower1ex\hbox{$\sim$}}}}
\def\gsim{\mathrel{\raise.3ex\hbox{$>$\kern-.75em\lower1ex\hbox{$\sim$}}}}
\def\ifmath#1{\relax\ifmmode #1\else $#1$\fi}

\def\simgt{\stackrel{>}{{}_\sim}}

\title{The Return of the WIMP: \\
Missing Energy Signals and the Galactic Center Excess}

\author[a,b]{Marcela~Carena,}
\author[c]{James Osborne,}
\author[c]{Nausheen~R.~Shah,}
\author[b,d]{Carlos~E.~M.~Wagner}

\affiliation[a]{Fermi National Accelerator Laboratory, P.~O.~Box 500, Batavia, IL 60510, USA}
\affiliation[b]{Enrico Fermi Institute and Kavli Institute for Cosmological Physics, University of Chicago, Chicago, IL 60637, USA}
\affiliation[c]{Department of Physics \& Astronomy, Wayne State University, Detroit, MI 48201, USA}
\affiliation[d]{HEP Division, Argonne National Laboratory, 9700 Cass Ave., Argonne, IL 60439, USA}

\emailAdd{carena@fnal.gov}
\emailAdd{jaosborne@wayne.edu}
\emailAdd{nausheen.shah@wayne.edu}
\emailAdd{cwagner@anl.gov}

\preprint{FERMILAB-PUB-19-199-T
\\\phantom{0} \hfill EFI-19-8
\\\phantom{0} \hfill WSU-HEP-1904}

\abstract{
In a recent work, we emphasized that an excess in tri-lepton events plus missing energy observed by the ATLAS experiment at the LHC could be interpreted as a signal of low energy supersymmetry. In such a scenario the lightest neutralino mass is approximately $m_\chi \simeq 60$~GeV and the direct Dark Matter detection cross section is naturally below the current bound.  In this work we present simple extensions of this  scenario that  lead to an explanation of the gamma ray excess at the center of the galaxy observed by Fermi-LAT, as well as the anti-proton excess observed by AMS-02.  These extensions include the addition of a small CP violating phase in the neutralino sector or the addition of a light CP-odd Higgs scalar. Our study is of special relevance in view of a recent analysis that casts  doubt on the previously accepted preference for mili-second pulsars as the origin of the galactic center excess. 
}

\begin{document}
\notoc
\maketitle
\flushbottom

\section{Introduction}
\label{sec:introduction}

Low energy supersymmetry is  undergoing heavy scrutiny from a large number of experimental probes, including missing energy signatures at the LHC, and direct and indirect  Dark Matter~(DM) searches.  Currently, collider searches for low energy supersymmetry~(SUSY)  have led to no conclusive evidence of its existence at the weak scale. Strongly interacting supersymmetric particles are constrained by LHC experiments to be above a scale of the order of a TeV~\cite{Aaboud:2017vwy,Sirunyan:2018vjp,Aaboud:2017hrg,Aaboud:2017aeu,Sirunyan:2018omt,Sirunyan:2018lul}, whereas bounds on weakly interacting particles are far less severe, and are of the order of a few hundred GeV.  

In a previous article~\cite{Carena:2018nlf} we interpreted a recently observed excess in tri-lepton plus missing energy events~\cite{Aaboud:2018sua} in terms of low energy SUSY. The pertinent LHC analysis was performed by the ATLAS collaboration using a newly developed Recursive Jigsaw Reconstruction (RJR) method~\cite{Jackson:2016mfb,Jackson:2017gcy}, which is well suited to study compressed spectra, and is sensitive to regions of parameters that may lead to no apparent signal events in conventional searches~\cite{Aaboud:2018jiw,Sirunyan:2018ubx}.  The reported excess was further supported by  analyses performed by  the GAMBIT collaboration, which argued that since  the bounds from conventional searches were obtained by simplified scenarios, they may not accurately capture the properties of  signals obtained in  more realistic models~\cite{Kvellestad:2018akf,Athron:2018vxy}. The preferred region of masses to explain the LHC excess was found to be for a mostly bino-like neutralino with mass of about 60~GeV, and a mostly Wino-like neutralino with a mass of about 160~GeV.  Obtaining the proper Dark Matter~(DM) relic density lead to the preference for Higgsinos with masses below 500~GeV, and values of $\tan\beta \simgt 10$. This region of parameters leads to a sizable contribution to the muon anomalous magnetic moment favored experimentally~\cite{Bennett:2006fi}, and is also consistent with the requirement of obtaining the correct Higgs mass in the Minimal Supersymmetric 
Standard Model~(MSSM)~\cite{Bahl:2017aev,Bagnaschi:2014rsa,Draper:2013oza,Lee:2015uza,Vega:2015fna}.

Indirect searches for DM have led to the observation of a significant gamma ray excess at the center of the galaxy~\cite{Hooper:2010mq,TheFermi-LAT:2015kwa}.  If interpreted in terms of DM annihilating  mainly into bottom-quark pairs, the cross section needed to explain such an excess is close to the one necessary to explain the relic density in a freeze-out scenario~\cite{Hooper:2011ti,Abazajian:2012pn,Daylan:2014rsa,Agrawal:2014oha}. This numerical fact has led to the expectation that the annihilation of a standard Weakly Interacting Massive Particle (WIMP) may be responsible for the observed galactic center excess (GCE). The initial excitement was subsequently contained by an analysis that demonstrated a preference of point-like sources~(PSs) to DM as a possible source of the GCE. This suggested some unresolved astrophysical objects, most likely mili-second pulsars, as the origin of the observed excess~\cite{Lee:2015fea,Bartels:2015aea}. However, a recent re-examination~\cite{Leane:2019xiy} of the analysis of Ref.~\cite{Lee:2015fea} has shown that this reported preference for PSs may be due to a mis-modeling of the real unknown distribution of PSs in the inner galaxy, and calls into question any inference from Ref.~\cite{Lee:2015fea} that the DM contribution to the GCE is small. Namely, it was found in Ref.~\cite{Leane:2019xiy} that even if one injects an  artificially enhanced simulated DM signal~(without any point-like sources) into the real Fermi data, 
the analysis pipeline of Ref.~\cite{Lee:2015fea} still misattributes this to PSs. Therefore the DM interpretation which was previously thought to be disfavored, can be again considered as the possible origin of the GCE.  

Intriguingly, DM masses of about 60~GeV  best fit  the GCE, and as mentioned above, are consistent with the ones that lead to an explanation of  the  LHC tri-lepton signatures.   Moreover, a recent excess in antiprotons observed by the AMS-02 experiment~\cite{Aguilar:2016kjl} seems to be consistent with DM annihilating into bottom-quark pairs, with a similar range of DM masses, and for annihilation cross sections consistent with the ones necessary to explain the GCE~\cite{Cui:2016ppb,Cuoco:2016eej,Cui:2018klo,Cuoco:2017okh,Cholis:2019ejx}. In view of all the above,  it seems highly relevant to investigate extensions of our previous work~\cite{Carena:2018nlf}  that, without changing the main results allowing for the interpretation of a possible LHC excess in the ATLAs  tri-lepton plus missing energy signal, could lead to a DM  explanation of the GCE~(as well as the anti-proton excess observed by AMS-02). In Sec.~\ref{Sec:Extensions} we will discuss  two such  extensions involving  the CP-violating MSSM~(CPVMSSM) and the Next to Minimal Supersymmetric Standard Model~(NMSSM), respectively. We reserve Sec.~\ref{Sec:Conc} for our Conclusions.

\section{Galactic Center Excess in Supersymmetry}\label{Sec:Extensions}

In our previous work~\cite{Carena:2018nlf} the proper DM relic density was obtained by resonant annihilation mediated by the Standard Model~(SM)-like Higgs boson $h$ for a lightest bino-like neutralino DM candidate  $\chi$ with mass $\sim$~60 GeV.  The relevant Lagrangian density is given by
\begin{equation}
{\cal{L}} = - g_{h \chi \chi} \chi \chi h - m_\chi \chi \chi + \textrm{h.c. }
\end{equation}
If the mass parameter and the coupling, $m_\chi$ and $g_{h \chi\chi}$, are real, as  in the CP-conserving MSSM,  the $\chi$ annihilation process will be $p$-wave suppressed and therefore no relevant indirect detection signal is  expected. A pseudoscalar coupling of the DM candidate to a mediator is required to generate an $s$-wave annihilation cross section.  Further, as has been explained in Ref.~\cite{Cheung:2014lqa}, even for the case of a dominantly $s$-wave annihilation cross section, due to thermal fluctuations, the matching of a resonantly enhanced annihilation cross section at the freeze-out temperature $T_F$~(relevant for the relic density) and $T=0$~(relevant for indirect detection) is non-trivial. In particular, in order to get the right indirect detection signal together with an observationally consistent relic density, there must be a particular relation between the mediator couplings to DM and the mediator couplings to  SM particles. Actually, it is known that an explanation of the GCE within the CP-conserving MSSM demands masses larger than 60~GeV~\cite{Caron:2015wda,Freese:2015ysa,Gherghetta:2015ysa}.

\subsection{CPVMSSM}

If $m_\chi$ is complex, then an $s$-wave contribution to the $h$ mediated annihilation cross section will be generated. This happens for instance for a bino DM candidate in the presence of a non-vanishing argument for the bino mass parameter $M_1$. In the particular case of $\arg[M_1] = \pi/2$, after absorbing the phase into a redefinition of $\chi$, $h$ will obtain a pseudoscalar coupling to the Majorana neutralino~(proportional to ($\chi\chi-\bar{\chi}\bar{\chi}$)), and the annihilation  cross section will be entirely $s$-wave. Hence, for a generic complex $m_\chi$ one expects  potentially relevant indirect detection signals. In particular, a suitable choice of the CP-violating phase  allows for coupling relations required for the consistency of the GCE as well as yielding a relic density prediction consistent with observations.

\begin{table}[!t]
  \centering
  \renewcommand{\arraystretch}{1.2}
  \begin{tabular}{l c | l c | l c| l c} 
    \hline
    Param. & Value & Param. & [GeV] & Param. & [GeV] & Param. & [GeV] \\
    \hline \hline
    $\arg[M_1]$ & $5.8^\circ$ & $\mu$ & -300 & $M_3$ & 3000 & $A_t$ & 2500  \\
    $\tan\beta$ & 20 & $M_1$ & 63.425 & $M_{\widetilde{L}}$ & 3000 & $A_b$ & 2500 \\
    $M_{H^\pm}$ & 1500 GeV & $M_2$ & -185 & $M_{\widetilde{Q}}$ & 3000 & $A_\tau$ & 1000 \\
    \cline{1-4} 
 \hline \hline
  \end{tabular}
  \caption{Benchmark values of CPVMSSM input parameters for \texttt{micrOMEGAs}. The squark and slepton soft masses are degenerate between generations and chiralities, and all unlisted $A$-terms are zero.}
  \label{T:CPV}
\end{table}

\begin{table}[!h]
  \centering
  \renewcommand{\arraystretch}{1.2}
  \begin{tabular}{l c | l c | l c | l c}
    \hline
    Part. & $m$ [GeV] & Part. & $m$ [GeV] & Part. & $m$ [GeV] & Part. & $m$ [GeV] \\
    \hline \hline
    $h$ & 125.5 & $\widetilde{\chi}_1^\pm$ & 165.2 & $\widetilde{\nu}_e$ & 2999.3 & $\widetilde{u}_R$ & 2999.8 \\
    $H_2$ & 1497.9 & $\widetilde{\chi}_2^\pm$ & 331.9 & $\widetilde{\nu}_\mu$ & 2999.3 & $\widetilde{u}_L$ & 2999.5 \\
    $H_3$ & 1497.9 & $\widetilde{\tau}_1$ & 2998.4 & $\widetilde{\nu}_{\tau}$ & 2999.3 & $\widetilde{d}_R$ & 3000.1 \\
    $H^\pm$ & 1500.0 & $\widetilde{\tau}_2$ & 3002.3 & $\widetilde{g}$ & 3000.0 & $\widetilde{d}_L$ & 3000.6 \\
    $\widetilde{\chi}_1^0$ & 62.7 & $\widetilde{e}_R$ & 3000.3 & $\widetilde{t}_1$ & 2945.8 & $\widetilde{s}_R$ & 3000.1 \\
    $\widetilde{\chi}_2^0$ & 165.0 & $\widetilde{e}_L$ & 3000.4 & $\widetilde{t}_2$ & 3058.4 & $\widetilde{s}_L$ & 3000.6 \\
    $\widetilde{\chi}_3^0$ & 309.6 & $\widetilde{\mu}_R$ & 3000.3 & $\widetilde{b}_1$ & 2997.6 & $\widetilde{c}_R$ & 2999.8 \\
    $\widetilde{\chi}_4^0$ & 329.0 & $\widetilde{\mu}_L$ & 3000.4 & $\widetilde{b}_2$ & 3003.1 & $\widetilde{c}_L$ & 2999.5 \\
    \hline
  \end{tabular}
  \caption{ Benchmark mass spectrum generated from the input parameters of Table~\ref{T:CPV}.}
  \label{T:CPV_spc}
\end{table}

Although the introduction of a non-vanishing phase to  $M_1$ does not impact the tri-lepton signatures at the LHC, it does lead to the appearance of electric dipole moments~(EDMs) which are strongly constrained experimentally~\cite{Andreev:2018ayy}.  The appearance of CP-violation only in the bino mass parameter, together with a sizable value of the heavy Higgs boson masses, $m_{H^+} > 1$~TeV, suppresses the two-loop contributions to the EDMs~\cite{Chang:2005ac,Giudice:2005rz,Ellis:2008zy,Li:2008ez}. In this case,  the main contribution to the electron EDM comes at 1-loop, and scales as the inverse square of the selectron masses. Values of  the selectron masses of  a few TeV are sufficient to evade the current EDM bounds. However, slepton masses at a  scale of a few~TeV will not give any significant contribution to the  anomalous magnetic moment of the muon, which then remains suppressed as compared to the experimentally favored value.

We numerically verified the above conclusions  by using the CPVMSSM implementation in \texttt{micrOMEGAs~5.0.8}~\cite{Belanger:2013oya,Belanger:2018mqt} which uses \texttt{CPsuperH~2.3}~\cite{Lee:2003nta,Lee:2012wa} as the spectrum generator.  We found that for
\begin{equation}
m_\chi = |m_\chi| \exp( i \phi),
\end{equation}
small values of $\phi$ for a neutralino mass $|m_\chi|\sim 60$ GeV can lead to resonant annihilation via $h$, and one may obtain consistency with the GCE signatures and the observed relic density. The full set of CPVMSSM parameters are shown in Table~\ref{T:CPV}, leading to approximately the same physical spectrum, tabulated in Table~\ref{T:CPV_spc}, as the one presented in Ref.~\cite{Carena:2018nlf}.  We take all SUSY scalar  masses of order $M_S\simeq 3$~TeV, which suppresses all EDMs, and the value of $A_t \simeq 2.5$~TeV, to obtain the right Higgs boson mass. Values of the other parameters are similar to the ones of the benchmark presented in Ref.~\cite{Carena:2018nlf}.

The small difference in the wino mass parameter, $M_2$, compared to the value presented in Ref.~\cite{Carena:2018nlf} is mostly due to the loop corrections to the neutralino and chargino spectrum present in the CP-conserving MSSM implementation of \texttt{micrOMEGAs} which was used in our previous work. The difference in $A_t$ is related to the different loop corrections used to compute the CP-even Higgs mass in both programs.  

Further choosing the heavy Higgs boson masses to be of about 1.5~TeV we obtain
\begin{alignat}{3}
 & \Omega h^2  = 0.119 , && \sigma^p_{\rm SI} = 2.17 \times 10^{-12}~{\rm pb} , && \sigma^n_{\rm SI} = 1.84 \times 10^{-12}~{\rm pb} , \nonumber \\
  & \sigma v |_{v = 0}  = 2.69  \times 10^{-26}~{\rm cm}^3/{\rm s} , \;\;\;\;\;\; && \sigma^p_{\rm SD} = 1.76 \times 10^{-5}~{\rm pb} , \;\;\;\;\;\; && \sigma^n_{\rm SD} = 1.36 \times 10^{-5}~{\rm pb} .
\end{alignat}
 In the above, $\Omega h^2$ is the DM relic density, $\sigma v |_{v = 0}$ is the annihilation cross section at $T=0$, and $\sigma^{n,p}_{\rm SI,SD}$ are the DM-neutron and -proton spin independent~(SI) and spin dependent~(SD) cross sections, respectively.  The above numbers are  in good agreement with current bounds~\cite{Aprile:2017iyp,Aprile:2018dbl,Akerib:2016vxi,Akerib:2017kat,Amole:2017dex,Amole:2019fdf}. 
 As we explained in Ref.~\cite{Carena:2018nlf}, similar results for the direct detection  cross section are obtained for other choices of the heavy Higgs masses of the order of the TeV scale. Moreover, since the annihilation cross section is mediated by the SM-like Higgs boson, the zero temperature effective annihilation cross section into $b\bar{b}$ final states is  $1.6 \times 10^{-26}~{\rm cm}^3/{\rm s}$, in good agreement with the necessary one to explain the GCE.  The remaining contributions to  $\sigma v |_{v = 0}$ from  other channels are approximately proportional to  the SM-like Higgs branching ratios (22\% into $WW^*$, 8\% into gluons, 7\% into tau leptons, etc).  For $m_\chi\sim60$ GeV, the main effect of these additional channels will be to slightly broaden the gamma ray spectrum without affecting the main contribution to the GCE in a relevant way. 
 
 For the chosen values of the parameters, which include vanishing trilinear slepton mass parameters, $A_l = 0$, the  electron EDM obtained with  \texttt{CPsuperH} is $(1.8 \times 10^{-30})$ e cm, which is approximately a factor five lower than the current bound on this quantity~\cite{Andreev:2018ayy}.  Actually, for these large values of the slepton masses there is a partial cancellation between the 1- and 2-loop contributions that suppresses the electron EDM from its 1-loop value~\cite{Ibrahim:2007fb}.  Indeed, whereas values of the electron EDM of the order of the current experimental bound are obtained for $m_{\tilde{L}} = 2$~TeV,  an approximate cancellation between the 1- and 2-loop contributions occurs for $m_{\tilde{L}} \sim 4$~TeV.

 \subsection{NMSSM}

  \begin{table}[!t]
  \centering
  \renewcommand{\arraystretch}{1.2}
  \begin{tabular}{l c | l c | l c | l c} 
    \hline
    Param. & Value & Param. & [GeV] & Param. & [GeV] & Param. & [GeV] \\
    \hline \hline
    $\tan \beta$ & 20 &$\mu_\textrm{eff}$ & -300 & $M_3$ & 3000  & $A_\lambda$ & -1260 \\
    $\lambda$ & 0.15 & $M_1$ & 62.62 & $M_{\widetilde{L}}$ & 450 & $A_\kappa$ & -10.8 \\
    $\kappa$ & -0.55 &  $M_2$ & -171. &  $M_{\widetilde{Q}}$ & 3000 & $A_t$ & 4000 \\
    \hline
  \end{tabular}
  \caption{Benchmark values of NMSSM input parameters for \texttt{micrOMEGAs}. The squark and slepton soft masses are degenerate between generations and chiralities, and all unlisted $A$-terms are set to $1$~TeV.}
  \label{T:NMSSM1}
\end{table}
\begin{table}[!h]
  \centering
  \renewcommand{\arraystretch}{1.2}
  \begin{tabular}{l c | l c | l c | l c}
    \hline
    Part. & $m$ [GeV] & Part. & $m$ [GeV] & Part. & $m$ [GeV] & Part. & $m$ [GeV] \\
    \hline \hline
    $h$ & 124.8 & $\widetilde{\chi}_1^\pm$ & 165.2 & $A_1$ & 120.8 & $\widetilde{u}_R$ & 3100.7 \\
    $H_2$ & 969.6 & $\widetilde{\chi}_2^\pm$ & 336.7 & $A_2$ & 974.1 & $\widetilde{u}_L$ & 3100.5 \\
    $H_3$ & 2185.5 & $\widetilde{\tau}_1$ & 438.3 & $\widetilde{\nu}_{e,\mu,\tau}$ & 445.7 & $\widetilde{d}_R$ & 3101.0 \\
    $H^\pm$ & 972.9 & $\widetilde{\tau}_2$ & 465.5 & $\widetilde{g}$ & 3198.1 & $\widetilde{d}_L$ & 3101.5 \\
    $\widetilde{\chi}_1^0$ & 60.7 & $\widetilde{e}_R$ & 452.0 & $\widetilde{t}_1$ & 2955.6 & $\widetilde{s}_R$ & 3101.0 \\
    $\widetilde{\chi}_2^0$ & 165.0 & $\widetilde{e}_L$ & 452.3 & $\widetilde{t}_2$ & 3120.5 & $\widetilde{s}_L$ & 3101.5 \\
    $\widetilde{\chi}_3^0$ & 315.8 & $\widetilde{\mu}_R$ & 452.0 & $\widetilde{b}_1$ & 3076.3 & $\widetilde{c}_R$ & 3100.7 \\
    $\widetilde{\chi}_4^0$ & 333.9 & $\widetilde{\mu}_L$ & 452.3 & $\widetilde{b}_2$ & 3077.8 & $\widetilde{c}_L$ & 3100.5 \\
    \hline
  \end{tabular}
  \caption{NMSSM Benchmark mass spectrum generated from the input parameters of Table~\ref{T:NMSSM1}.}
  \label{T:NMSSM_spc}
\end{table}

An alternative scenario, which was advocated in the context of the GCE in Ref.~\cite{Cheung:2014lqa}, is the  NMSSM. The appearance of additional CP-odd and CP-even singlets in the NMSSM allows for the presence of extra channels contributing to the  resonant annihilation of  DM. For a 60 GeV, mostly bino DM candidate, compatibility with the GCE and relic density may be obtained via resonant annihilation mediated by both the  CP-odd singlet  $A_1$~(to give the relevant $s$-wave contribution at zero temperature) as well as through $h$~(to provide the dominant contribution to the finite temperature annihilation cross section relevant for the relic density).\footnote{ Details of such a mechanism may be found in Ref.~\cite{Cheung:2014lqa}, where it is shown that for a slightly lighter singlino-Higgsino DM candidate of mass $m_\chi\sim 40$ GeV, $A_1$ can assist annihilation close to the $Z$ funnel, and yield   consistency with the GCE and the observed relic density.}  

One advantage of the NMSSM compared to the CPVMSSM scenario is the possibility of preserving CP and hence avoiding the EDM constraints. One can then lower the slepton masses to values of order 400~GeV, leading to a sizable contribution to the muon anomalous magnetic moment, which is experimentally favored~\cite{Bennett:2006fi}. Using \texttt{NMSSMTools 5.1.1}~\cite{Belanger:2005kh, Domingo:2007dx, Djouadi:1997yw,Degrassi:2009yq, Ellwanger:2005dv, Ellwanger:2004xm} which is the spectrum generator for the NMSSM in \texttt{micrOMEGAs~5.0.8}, we choose  parameters tabulated in Table~\ref{T:NMSSM1}, such that, as  shown in Table~\ref{T:NMSSM_spc}, a similar neutralino/chargino spectrum as in our previous work~\cite{Carena:2018nlf} is obtained. These parameters are consistent with the observed ATLAS excess. Moreover, with the choice of slepton masses  $M_{\tilde{L}}=450$ GeV, these parameter choices also lead to consistency with the muon anomalous magnetic moment, $a_\mu^{\rm MSSM} = 217\times 10^{-11}$. 

The values of the parameters in the singlet sector are selected to keep the lightest neutralino MSSM-like.  This is achieved for $\left | \kappa/\lambda \right | >1$, for which the singlino state becomes heavier than the Higgsinos. Since the singlet CP-even scalar tends to be heavy in this region of parameters it does not lead to any large mixing effects on the SM-like Higgs properties. The value of $A_\kappa$ is chosen to obtain a lightest CP-odd mass consistent with the resonant annihilation condition.  Finally, the value of $A_\lambda$ was chosen to ensure a heavy enough doublet Higgs spectrum to avoid direct LHC search constraints, and to provide significant cancellations in the direct detection cross section amplitudes~\cite{Huang:2014xua,Cheung:2014lqa}.  Observe that the same physical spectrum and hence the same physical results  can be  obtained for a broad range of correlated values of the model parameters. 

One obtains a lightest neutralino mass of order of 60~GeV, a second lightest neutralino of about 165~GeV, and a light pseudoscalar with mass $\sim 120$ GeV. The doublet-like Higgs boson masses are $m_{A_2}\sim m_{H_2}\sim 970$ GeV, and the heavier CP-even singlet is   $\sim2.2$ TeV. With this mass spectrum, one obtains
\begin{alignat}{3}
  & \Omega h^2 = 0.119 , && \sigma^p_{\rm SI} = 5.6 \times 10^{-12}~{\rm pb} , && \sigma^n_{\rm SI} = 7.23 \times 10^{-12}~{\rm pb} , \nonumber \\
  & \sigma v |_{v = 0} = 2.25 \times 10^{-26}~{\rm cm}^3/{\rm s} , \;\;\;\;\;\; && \sigma^p_{\rm SD} = 1.59 \times 10^{-5}~{\rm pb} , \;\;\;\;\;\; && \sigma^n_{\rm SD} = 1.23 \times 10^{-5}~{\rm pb} .
\end{alignat}
In this scenario, since the zero temperature annihilation cross section is mediated primarily by the light pseudoscalar $A_1$, the contribution of the $b\bar{b}$ channel to $\sigma v |_{v = 0}$ is  about 90\%. Hence, again, the cross section is consistent with the one necessary to explain the GCE. As in the previous scenario, the SI and SD direct detection cross sections are in good agreement with current bounds~\cite{Aprile:2017iyp,Aprile:2018dbl,Akerib:2016vxi,Akerib:2017kat,Amole:2017dex,Amole:2019fdf}.

Observe that, albeit small in the resonant case discussed, the mixing of the singlet pseudoscalar and the doublet pseudoscalar state is crucial for obtaining the GCE. In principle, such a mixing also leads  to an increase of the production cross section of $A_1$ at colliders which can be searched for at the LHC. However, even though the lightest pseudoscalar has a mass of about 120~GeV, with $BR(A_1 \to b\bar{b}) \sim 90$\% and $BR(A_1 \sim \tau^+\tau^-) \sim 10$\%, respectively,  it's effective gluon-fusion production cross section is only  $\mathcal{O}$(1)~pb. Such a small production cross section  makes it challenging to test at the LHC. The prospects for the heavy  CP-even and CP-odd doublet-like Higgs bosons are more interesting. On the one hand, the decays of both $H_2$ and $A_2$ into many different weakly interacting states (such as charginos, neutralinos and staus) suppresses their $\tau$-pair signatures, allowing them to evade current limits. On the other hand, their branching ratio into $\tau$-pairs is still $\sim$4\%, which makes them potentially detectable in this channel with more luminosity. Further, their decays into charginos/neutralinos as well as cascade decays into additional Higgs bosons may provide complimentary search handles at the  LHC~\cite{Baum:2017gbj,Baum:2018zhf, Baum:2019uzg, Baum:2019pqc, Gori:2018pmk, Adhikary:2018ise, Basler:2018dac}.

\section{Conclusions}\label{Sec:Conc}

In this work, we have presented scenarios that lead to a simultaneous explanation of an excess in tri-leptons plus missing energy at the LHC, a gamma-ray excess at the center of the galaxy, and an antiproton excess in cosmic rays.  These are based on either the MSSM with CP-violation in the neutralino sector, or on the NMSSM with a light CP-odd scalar.  The lightest neutralino acquires a mass of order 60~GeV, while the second lightest neutralino and the lightest chargino have masses of order 165~GeV.   
While the CPVMSSM scenario may lead to an observable electron EDM,
the NMSSM scenario may be tested by searches at the LHC for heavy scalars decaying to  $\tau$ pairs, electroweakinos or additional scalars. The latter scenario also gives rise to values of the muon anomalous magnetic moment consistent with current observations. 

In closing let us emphasize  that while the chargino and neutralino spectrum leading to the explanation of the LHC tri-lepton excess depends crucially on $M_2$, the GCE  is fairly independent of the value of $M_2$.  Hence, if the LHC excess were not confirmed, the same scenarios, but for somewhat larger values of $M_2$, would still lead to an explanation of the galactic center and antiproton excess, consistent with all observational constraints.

\section*{\sc Acknowledgments}

We thank D. Hooper for initially suggesting the idea of  CP-violation in the DM-Higgs coupling. We additionally thank R. Leanne and T. Slatyer for useful discussions and comments. This manuscript has been authored by Fermi Research Alliance, LLC under Contract No. DE-AC02-07CH11359 with the U.S. Department of Energy, Office of Science, Office of High Energy Physics.  Work at University of Chicago is supported in part by U.S. Department of Energy grant number DE-FG02-13ER41958. Work at ANL is supported in part by the U.S. Department of Energy under Contract No. DE-AC02-06CH11357. NRS and JO are supported by Wayne State University and by the U.S. Department of Energy under Contract No. DESC0007983.

\bibliography{mybibfile}

\providecommand{\href}[2]{#2}\begingroup\raggedright\begin{thebibliography}{10}

\bibitem{Aaboud:2017vwy}
{\scshape ATLAS} collaboration, M.~Aaboud et~al., \emph{{Search for squarks and
  gluinos in final states with jets and missing transverse momentum using
  36~fb$^{-1}$ of $\sqrt{s}=13$~TeV pp collision data with the ATLAS
  detector}}, \href{http://dx.doi.org/10.1103/PhysRevD.97.112001}{\emph{Phys.
  Rev.} {\bf D97} (2018) 112001}, [\href{https://arxiv.org/abs/1712.02332}{{\tt
  1712.02332}}].

\bibitem{Sirunyan:2018vjp}
{\scshape CMS} collaboration, A.~M. Sirunyan et~al., \emph{{Search for natural
  and split supersymmetry in proton-proton collisions at $ \sqrt{s}=13 $ TeV in
  final states with jets and missing transverse momentum}},
  \href{http://dx.doi.org/10.1007/JHEP05(2018)025}{\emph{JHEP} {\bf 05} (2018)
  025}, [\href{https://arxiv.org/abs/1802.02110}{{\tt 1802.02110}}].

\bibitem{Aaboud:2017hrg}
{\scshape ATLAS} collaboration, M.~Aaboud et~al., \emph{{Search for
  supersymmetry in final states with missing transverse momentum and multiple
  $b$-jets in proton-proton collisions at $ \sqrt{s}=13 $ TeV with the ATLAS
  detector}}, \href{http://dx.doi.org/10.1007/JHEP06(2018)107}{\emph{JHEP} {\bf
  06} (2018) 107}, [\href{https://arxiv.org/abs/1711.01901}{{\tt 1711.01901}}].

\bibitem{Aaboud:2017aeu}
{\scshape ATLAS} collaboration, M.~Aaboud et~al., \emph{{Search for top-squark
  pair production in final states with one lepton, jets, and missing transverse
  momentum using 36 fb$^{-1}$ of $ \sqrt{s}=13 $ TeV pp collision data with the
  ATLAS detector}},
  \href{http://dx.doi.org/10.1007/JHEP06(2018)108}{\emph{JHEP} {\bf 06} (2018)
  108}, [\href{https://arxiv.org/abs/1711.11520}{{\tt 1711.11520}}].

\bibitem{Sirunyan:2018omt}
{\scshape CMS} collaboration, A.~M. Sirunyan et~al., \emph{{Search for top
  squarks decaying via four-body or chargino-mediated modes in single-lepton
  final states in proton-proton collisions at $\sqrt{s} =$ 13 TeV}},
  \href{https://arxiv.org/abs/1805.05784}{{\tt 1805.05784}}.

\bibitem{Sirunyan:2018lul}
{\scshape CMS} collaboration, A.~M. Sirunyan et~al., \emph{{Searches for pair
  production of charginos and top squarks in final states with two oppositely
  charged leptons in proton-proton collisions at $\sqrt{s}=$ 13 TeV}},
  {\emph{Submitted to: JHEP} (2018) },
  [\href{https://arxiv.org/abs/1807.07799}{{\tt 1807.07799}}].

\bibitem{Carena:2018nlf}
M.~Carena, J.~Osborne, N.~R. Shah and C.~E.~M. Wagner, \emph{{Supersymmetry and
  LHC Missing Energy Signals}},
  \href{http://dx.doi.org/10.1103/PhysRevD.98.115010}{\emph{Phys. Rev.} {\bf
  D98} (2018) 115010}, [\href{https://arxiv.org/abs/1809.11082}{{\tt
  1809.11082}}].

\bibitem{Aaboud:2018sua}
{\scshape ATLAS} collaboration, M.~Aaboud et~al., \emph{{Search for
  chargino-neutralino production using recursive jigsaw reconstruction in final
  states with two or three charged leptons in proton-proton collisions at
  $\sqrt{s}=13$ TeV with the ATLAS detector}},
  \href{https://arxiv.org/abs/1806.02293}{{\tt 1806.02293}}.

\bibitem{Jackson:2016mfb}
P.~Jackson, C.~Rogan and M.~Santoni, \emph{{Sparticles in motion: Analyzing
  compressed SUSY scenarios with a new method of event reconstruction}},
  \href{http://dx.doi.org/10.1103/PhysRevD.95.035031}{\emph{Phys. Rev.} {\bf
  D95} (2017) 035031}, [\href{https://arxiv.org/abs/1607.08307}{{\tt
  1607.08307}}].

\bibitem{Jackson:2017gcy}
P.~Jackson and C.~Rogan, \emph{{Recursive Jigsaw Reconstruction: HEP event
  analysis in the presence of kinematic and combinatoric ambiguities}},
  \href{http://dx.doi.org/10.1103/PhysRevD.96.112007}{\emph{Phys. Rev.} {\bf
  D96} (2017) 112007}, [\href{https://arxiv.org/abs/1705.10733}{{\tt
  1705.10733}}].

\bibitem{Aaboud:2018jiw}
{\scshape ATLAS} collaboration, M.~Aaboud et~al., \emph{{Search for electroweak
  production of supersymmetric particles in final states with two or three
  leptons at $\sqrt{s}=13\,$TeV with the ATLAS detector}},
  \href{https://arxiv.org/abs/1803.02762}{{\tt 1803.02762}}.

\bibitem{Sirunyan:2018ubx}
{\scshape CMS} collaboration, A.~M. Sirunyan et~al., \emph{{Combined search for
  electroweak production of charginos and neutralinos in proton-proton
  collisions at $\sqrt{s} =$ 13 TeV}},
  \href{http://dx.doi.org/10.1007/JHEP03(2018)160}{\emph{JHEP} {\bf 03} (2018)
  160}, [\href{https://arxiv.org/abs/1801.03957}{{\tt 1801.03957}}].

\bibitem{Kvellestad:2018akf}
{\scshape GAMBIT} collaboration, A.~Kvellestad, \emph{{Global fits of the MSSM
  with GAMBIT}},  in \emph{{39th International Conference on High Energy
  Physics (ICHEP 2018) Seoul, Gangnam-Gu, Korea, Republic of, July 4-11,
  2018}}, 2018.
\newblock \href{https://arxiv.org/abs/1807.03208}{{\tt 1807.03208}}.

\bibitem{Athron:2018vxy}
{\scshape GAMBIT} collaboration, P.~Athron et~al., \emph{{Combined collider
  constraints on neutralinos and charginos}},
  \href{https://arxiv.org/abs/1809.02097}{{\tt 1809.02097}}.

\bibitem{Bennett:2006fi}
{\scshape Muon g-2} collaboration, G.~W. Bennett et~al., \emph{{Final Report of
  the Muon E821 Anomalous Magnetic Moment Measurement at BNL}},
  \href{http://dx.doi.org/10.1103/PhysRevD.73.072003}{\emph{Phys. Rev.} {\bf
  D73} (2006) 072003}, [\href{https://arxiv.org/abs/hep-ex/0602035}{{\tt
  hep-ex/0602035}}].

\bibitem{Bahl:2017aev}
H.~Bahl, S.~Heinemeyer, W.~Hollik and G.~Weiglein, \emph{{Reconciling EFT and
  hybrid calculations of the light MSSM Higgs-boson mass}},
  \href{http://dx.doi.org/10.1140/epjc/s10052-018-5544-3}{\emph{Eur. Phys. J.}
  {\bf C78} (2018) 57}, [\href{https://arxiv.org/abs/1706.00346}{{\tt
  1706.00346}}].

\bibitem{Bagnaschi:2014rsa}
E.~Bagnaschi, G.~F. Giudice, P.~Slavich and A.~Strumia, \emph{{Higgs Mass and
  Unnatural Supersymmetry}},
  \href{http://dx.doi.org/10.1007/JHEP09(2014)092}{\emph{JHEP} {\bf 09} (2014)
  092}, [\href{https://arxiv.org/abs/1407.4081}{{\tt 1407.4081}}].

\bibitem{Draper:2013oza}
P.~Draper, G.~Lee and C.~E.~M. Wagner, \emph{{Precise estimates of the Higgs
  mass in heavy supersymmetry}},
  \href{http://dx.doi.org/10.1103/PhysRevD.89.055023}{\emph{Phys. Rev.} {\bf
  D89} (2014) 055023}, [\href{https://arxiv.org/abs/1312.5743}{{\tt
  1312.5743}}].

\bibitem{Lee:2015uza}
G.~Lee and C.~E.~M. Wagner, \emph{{Higgs bosons in heavy supersymmetry with an
  intermediate m$_A$}},
  \href{http://dx.doi.org/10.1103/PhysRevD.92.075032}{\emph{Phys. Rev.} {\bf
  D92} (2015) 075032}, [\href{https://arxiv.org/abs/1508.00576}{{\tt
  1508.00576}}].

\bibitem{Vega:2015fna}
J.~Pardo~Vega and G.~Villadoro, \emph{{SusyHD: Higgs mass Determination in
  Supersymmetry}}, \href{http://dx.doi.org/10.1007/JHEP07(2015)159}{\emph{JHEP}
  {\bf 07} (2015) 159}, [\href{https://arxiv.org/abs/1504.05200}{{\tt
  1504.05200}}].

\bibitem{Hooper:2010mq}
D.~Hooper and L.~Goodenough, \emph{{Dark Matter Annihilation in The Galactic
  Center As Seen by the Fermi Gamma Ray Space Telescope}},
  \href{http://dx.doi.org/10.1016/j.physletb.2011.02.029}{\emph{Phys. Lett.}
  {\bf B697} (2011) 412--428}, [\href{https://arxiv.org/abs/1010.2752}{{\tt
  1010.2752}}].

\bibitem{TheFermi-LAT:2015kwa}
{\scshape Fermi-LAT} collaboration, M.~Ajello et~al., \emph{{Fermi-LAT
  Observations of High-Energy $\gamma$-Ray Emission Toward the Galactic
  Center}},
  \href{http://dx.doi.org/10.3847/0004-637X/819/1/44}{\emph{Astrophys. J.} {\bf
  819} (2016) 44}, [\href{https://arxiv.org/abs/1511.02938}{{\tt 1511.02938}}].

\bibitem{Hooper:2011ti}
D.~Hooper and T.~Linden, \emph{{On The Origin Of The Gamma Rays From The
  Galactic Center}},
  \href{http://dx.doi.org/10.1103/PhysRevD.84.123005}{\emph{Phys. Rev.} {\bf
  D84} (2011) 123005}, [\href{https://arxiv.org/abs/1110.0006}{{\tt
  1110.0006}}].

\bibitem{Abazajian:2012pn}
K.~N. Abazajian and M.~Kaplinghat, \emph{{Detection of a Gamma-Ray Source in
  the Galactic Center Consistent with Extended Emission from Dark Matter
  Annihilation and Concentrated Astrophysical Emission}},
  \href{http://dx.doi.org/10.1103/PhysRevD.86.083511,
  10.1103/PhysRevD.87.129902}{\emph{Phys. Rev.} {\bf D86} (2012) 083511},
  [\href{https://arxiv.org/abs/1207.6047}{{\tt 1207.6047}}].

\bibitem{Daylan:2014rsa}
T.~Daylan, D.~P. Finkbeiner, D.~Hooper, T.~Linden, S.~K.~N. Portillo, N.~L.
  Rodd et~al., \emph{{The characterization of the gamma-ray signal from the
  central Milky Way: A case for annihilating dark matter}},
  \href{http://dx.doi.org/10.1016/j.dark.2015.12.005}{\emph{Phys. Dark Univ.}
  {\bf 12} (2016) 1--23}, [\href{https://arxiv.org/abs/1402.6703}{{\tt
  1402.6703}}].

\bibitem{Agrawal:2014oha}
P.~Agrawal, B.~Batell, P.~J. Fox and R.~Harnik, \emph{{WIMPs at the Galactic
  Center}}, \href{http://dx.doi.org/10.1088/1475-7516/2015/05/011}{\emph{JCAP}
  {\bf 1505} (2015) 011}, [\href{https://arxiv.org/abs/1411.2592}{{\tt
  1411.2592}}].

\bibitem{Lee:2015fea}
S.~K. Lee, M.~Lisanti, B.~R. Safdi, T.~R. Slatyer and W.~Xue, \emph{{Evidence
  for Unresolved $\gamma$-Ray Point Sources in the Inner Galaxy}},
  \href{http://dx.doi.org/10.1103/PhysRevLett.116.051103}{\emph{Phys. Rev.
  Lett.} {\bf 116} (2016) 051103},
  [\href{https://arxiv.org/abs/1506.05124}{{\tt 1506.05124}}].

\bibitem{Bartels:2015aea}
R.~Bartels, S.~Krishnamurthy and C.~Weniger, \emph{{Strong support for the
  millisecond pulsar origin of the Galactic center GeV excess}},
  \href{http://dx.doi.org/10.1103/PhysRevLett.116.051102}{\emph{Phys. Rev.
  Lett.} {\bf 116} (2016) 051102},
  [\href{https://arxiv.org/abs/1506.05104}{{\tt 1506.05104}}].

\bibitem{Leane:2019xiy}
R.~K. Leane and T.~R. Slatyer, \emph{{Dark Matter Strikes Back at the Galactic
  Center}},  \href{https://arxiv.org/abs/1904.08430}{{\tt 1904.08430}}.

\bibitem{Aguilar:2016kjl}
{\scshape AMS} collaboration, M.~Aguilar et~al., \emph{{Antiproton Flux,
  Antiproton-to-Proton Flux Ratio, and Properties of Elementary Particle Fluxes
  in Primary Cosmic Rays Measured with the Alpha Magnetic Spectrometer on the
  International Space Station}},
  \href{http://dx.doi.org/10.1103/PhysRevLett.117.091103}{\emph{Phys. Rev.
  Lett.} {\bf 117} (2016) 091103}.

\bibitem{Cui:2016ppb}
M.-Y. Cui, Q.~Yuan, Y.-L.~S. Tsai and Y.-Z. Fan, \emph{{Possible dark matter
  annihilation signal in the AMS-02 antiproton data}},
  \href{http://dx.doi.org/10.1103/PhysRevLett.118.191101}{\emph{Phys. Rev.
  Lett.} {\bf 118} (2017) 191101},
  [\href{https://arxiv.org/abs/1610.03840}{{\tt 1610.03840}}].

\bibitem{Cuoco:2016eej}
A.~Cuoco, M.~Krämer and M.~Korsmeier, \emph{{Novel Dark Matter Constraints from
  Antiprotons in Light of AMS-02}},
  \href{http://dx.doi.org/10.1103/PhysRevLett.118.191102}{\emph{Phys. Rev.
  Lett.} {\bf 118} (2017) 191102},
  [\href{https://arxiv.org/abs/1610.03071}{{\tt 1610.03071}}].

\bibitem{Cui:2018klo}
M.-Y. Cui, X.~Pan, Q.~Yuan, Y.-Z. Fan and H.-S. Zong, \emph{{Revisit of cosmic
  ray antiprotons from dark matter annihilation with updated constraints on the
  background model from AMS-02 and collider data}},
  \href{http://dx.doi.org/10.1088/1475-7516/2018/06/024}{\emph{JCAP} {\bf 1806}
  (2018) 024}, [\href{https://arxiv.org/abs/1803.02163}{{\tt 1803.02163}}].

\bibitem{Cuoco:2017okh}
A.~Cuoco, J.~Heisig, M.~Korsmeier and M.~Krämer, \emph{{A combined dark matter
  study of AMS-02 antiprotons and Fermi-LAT gamma rays}},
  \href{http://dx.doi.org/10.22323/1.314.0065}{\emph{PoS} {\bf EPS-HEP2017}
  (2017) 065}, [\href{https://arxiv.org/abs/1711.06460}{{\tt 1711.06460}}].

\bibitem{Cholis:2019ejx}
I.~Cholis, T.~Linden and D.~Hooper, \emph{{A Robust Excess in the Cosmic-Ray
  Antiproton Spectrum: Implications for Annihilating Dark Matter}},
  \href{https://arxiv.org/abs/1903.02549}{{\tt 1903.02549}}.

\bibitem{Cheung:2014lqa}
C.~Cheung, M.~Papucci, D.~Sanford, N.~R. Shah and K.~M. Zurek, \emph{{NMSSM
  Interpretation of the Galactic Center Excess}},
  \href{http://dx.doi.org/10.1103/PhysRevD.90.075011}{\emph{Phys. Rev.} {\bf
  D90} (2014) 075011}, [\href{https://arxiv.org/abs/1406.6372}{{\tt
  1406.6372}}].

\bibitem{Caron:2015wda}
A.~Achterberg, S.~Amoroso, S.~Caron, L.~Hendriks, R.~Ruiz~de Austri and
  C.~Weniger, \emph{{A description of the Galactic Center excess in the Minimal
  Supersymmetric Standard Model}},
  \href{http://dx.doi.org/10.1088/1475-7516/2015/08/006}{\emph{JCAP} {\bf 1508}
  (2015) 006}, [\href{https://arxiv.org/abs/1502.05703}{{\tt 1502.05703}}].

\bibitem{Freese:2015ysa}
K.~Freese, A.~Lopez, N.~R. Shah and B.~Shakya, \emph{{MSSM A-funnel and the
  Galactic Center Excess: Prospects for the LHC and Direct Detection
  Experiments}}, \href{http://dx.doi.org/10.1007/JHEP04(2016)059}{\emph{JHEP}
  {\bf 04} (2016) 059}, [\href{https://arxiv.org/abs/1509.05076}{{\tt
  1509.05076}}].

\bibitem{Gherghetta:2015ysa}
T.~Gherghetta, B.~von Harling, A.~D. Medina, M.~A. Schmidt and T.~Trott,
  \emph{{SUSY implications from WIMP annihilation into scalars at the Galactic
  Center}}, \href{http://dx.doi.org/10.1103/PhysRevD.91.105004}{\emph{Phys.
  Rev.} {\bf D91} (2015) 105004}, [\href{https://arxiv.org/abs/1502.07173}{{\tt
  1502.07173}}].

\bibitem{Andreev:2018ayy}
{\scshape ACME} collaboration, V.~Andreev et~al., \emph{{Improved limit on the
  electric dipole moment of the electron}},
  \href{http://dx.doi.org/10.1038/s41586-018-0599-8}{\emph{Nature} {\bf 562}
  (2018) 355--360}.

\bibitem{Chang:2005ac}
D.~Chang, W.-F. Chang and W.-Y. Keung, \emph{{Electric dipole moment in the
  split supersymmetry models}},
  \href{http://dx.doi.org/10.1103/PhysRevD.71.076006}{\emph{Phys. Rev.} {\bf
  D71} (2005) 076006}, [\href{https://arxiv.org/abs/hep-ph/0503055}{{\tt
  hep-ph/0503055}}].

\bibitem{Giudice:2005rz}
G.~F. Giudice and A.~Romanino, \emph{{Electric dipole moments in split
  supersymmetry}},
  \href{http://dx.doi.org/10.1016/j.physletb.2006.01.027}{\emph{Phys. Lett.}
  {\bf B634} (2006) 307--314},
  [\href{https://arxiv.org/abs/hep-ph/0510197}{{\tt hep-ph/0510197}}].

\bibitem{Ellis:2008zy}
J.~R. Ellis, J.~S. Lee and A.~Pilaftsis, \emph{{Electric Dipole Moments in the
  MSSM Reloaded}},
  \href{http://dx.doi.org/10.1088/1126-6708/2008/10/049}{\emph{JHEP} {\bf 10}
  (2008) 049}, [\href{https://arxiv.org/abs/0808.1819}{{\tt 0808.1819}}].

\bibitem{Li:2008ez}
Y.~Li, S.~Profumo and M.~Ramsey-Musolf, \emph{{Bino-driven Electroweak
  Baryogenesis with highly suppressed Electric Dipole Moments}},
  \href{http://dx.doi.org/10.1016/j.physletb.2009.02.004}{\emph{Phys. Lett.}
  {\bf B673} (2009) 95--100}, [\href{https://arxiv.org/abs/0811.1987}{{\tt
  0811.1987}}].

\bibitem{Belanger:2013oya}
G.~Belanger, F.~Boudjema, A.~Pukhov and A.~Semenov, \emph{{micrOMEGAs3: A
  program for calculating dark matter observables}},
  \href{http://dx.doi.org/10.1016/j.cpc.2013.10.016}{\emph{Comput. Phys.
  Commun.} {\bf 185} (2014) 960--985},
  [\href{https://arxiv.org/abs/1305.0237}{{\tt 1305.0237}}].

\bibitem{Belanger:2018mqt}
G.~Bélanger, F.~Boudjema, A.~Goudelis, A.~Pukhov and B.~Zaldivar,
  \emph{{micrOMEGAs5.0 : Freeze-in}},
  \href{http://dx.doi.org/10.1016/j.cpc.2018.04.027}{\emph{Comput. Phys.
  Commun.} {\bf 231} (2018) 173--186},
  [\href{https://arxiv.org/abs/1801.03509}{{\tt 1801.03509}}].

\bibitem{Lee:2003nta}
J.~S. Lee, A.~Pilaftsis, M.~Carena, S.~Y. Choi, M.~Drees, J.~R. Ellis et~al.,
  \emph{{CPsuperH: A Computational tool for Higgs phenomenology in the minimal
  supersymmetric standard model with explicit CP violation}},
  \href{http://dx.doi.org/10.1016/S0010-4655(03)00463-6}{\emph{Comput. Phys.
  Commun.} {\bf 156} (2004) 283--317},
  [\href{https://arxiv.org/abs/hep-ph/0307377}{{\tt hep-ph/0307377}}].

\bibitem{Lee:2012wa}
J.~S. Lee, M.~Carena, J.~Ellis, A.~Pilaftsis and C.~E.~M. Wagner,
  \emph{{CPsuperH2.3: an Updated Tool for Phenomenology in the MSSM with
  Explicit CP Violation}},
  \href{http://dx.doi.org/10.1016/j.cpc.2012.11.006}{\emph{Comput. Phys.
  Commun.} {\bf 184} (2013) 1220--1233},
  [\href{https://arxiv.org/abs/1208.2212}{{\tt 1208.2212}}].

\bibitem{Aprile:2017iyp}
{\scshape XENON} collaboration, E.~Aprile et~al., \emph{{First Dark Matter
  Search Results from the XENON1T Experiment}},
  \href{http://dx.doi.org/10.1103/PhysRevLett.119.181301}{\emph{Phys. Rev.
  Lett.} {\bf 119} (2017) 181301},
  [\href{https://arxiv.org/abs/1705.06655}{{\tt 1705.06655}}].

\bibitem{Aprile:2018dbl}
{\scshape XENON} collaboration, E.~Aprile et~al., \emph{{Dark Matter Search
  Results from a One Tonne$\times$Year Exposure of XENON1T}},
  \href{https://arxiv.org/abs/1805.12562}{{\tt 1805.12562}}.

\bibitem{Akerib:2016vxi}
{\scshape LUX} collaboration, D.~S. Akerib et~al., \emph{{Results from a search
  for dark matter in the complete LUX exposure}},
  \href{http://dx.doi.org/10.1103/PhysRevLett.118.021303}{\emph{Phys. Rev.
  Lett.} {\bf 118} (2017) 021303},
  [\href{https://arxiv.org/abs/1608.07648}{{\tt 1608.07648}}].

\bibitem{Akerib:2017kat}
{\scshape LUX} collaboration, D.~S. Akerib et~al., \emph{{Limits on
  spin-dependent WIMP-nucleon cross section obtained from the complete LUX
  exposure}},
  \href{http://dx.doi.org/10.1103/PhysRevLett.118.251302}{\emph{Phys. Rev.
  Lett.} {\bf 118} (2017) 251302},
  [\href{https://arxiv.org/abs/1705.03380}{{\tt 1705.03380}}].

\bibitem{Amole:2017dex}
{\scshape PICO} collaboration, C.~Amole et~al., \emph{{Dark Matter Search
  Results from the PICO-60 C$_3$F$_8$ Bubble Chamber}},
  \href{http://dx.doi.org/10.1103/PhysRevLett.118.251301}{\emph{Phys. Rev.
  Lett.} {\bf 118} (2017) 251301},
  [\href{https://arxiv.org/abs/1702.07666}{{\tt 1702.07666}}].

\bibitem{Amole:2019fdf}
{\scshape PICO} collaboration, C.~Amole et~al., \emph{{Dark Matter Search
  Results from the Complete Exposure of the PICO-60 C$_3$F$_8$ Bubble
  Chamber}},  \href{https://arxiv.org/abs/1902.04031}{{\tt 1902.04031}}.

\bibitem{Ibrahim:2007fb}
T.~Ibrahim and P.~Nath, \emph{{CP Violation From Standard Model to Strings}},
  \href{http://dx.doi.org/10.1103/RevModPhys.80.577}{\emph{Rev. Mod. Phys.}
  {\bf 80} (2008) 577--631}, [\href{https://arxiv.org/abs/0705.2008}{{\tt
  0705.2008}}].

\bibitem{Belanger:2005kh}
G.~Belanger, F.~Boudjema, C.~Hugonie, A.~Pukhov and A.~Semenov, \emph{{Relic
  density of dark matter in the NMSSM}},
  \href{http://dx.doi.org/10.1088/1475-7516/2005/09/001}{\emph{JCAP} {\bf 0509}
  (2005) 001}, [\href{https://arxiv.org/abs/hep-ph/0505142}{{\tt
  hep-ph/0505142}}].

\bibitem{Domingo:2007dx}
F.~Domingo and U.~Ellwanger, \emph{{Updated Constraints from $B$ Physics on the
  MSSM and the NMSSM}},
  \href{http://dx.doi.org/10.1088/1126-6708/2007/12/090}{\emph{JHEP} {\bf 12}
  (2007) 090}, [\href{https://arxiv.org/abs/0710.3714}{{\tt 0710.3714}}].

\bibitem{Djouadi:1997yw}
A.~Djouadi, J.~Kalinowski and M.~Spira, \emph{{HDECAY: A Program for Higgs
  boson decays in the standard model and its supersymmetric extension}},
  \href{http://dx.doi.org/10.1016/S0010-4655(97)00123-9}{\emph{Comput. Phys.
  Commun.} {\bf 108} (1998) 56--74},
  [\href{https://arxiv.org/abs/hep-ph/9704448}{{\tt hep-ph/9704448}}].

\bibitem{Degrassi:2009yq}
G.~Degrassi and P.~Slavich, \emph{{On the radiative corrections to the neutral
  Higgs boson masses in the NMSSM}},
  \href{http://dx.doi.org/10.1016/j.nuclphysb.2009.09.018}{\emph{Nucl. Phys.}
  {\bf B825} (2010) 119--150}, [\href{https://arxiv.org/abs/0907.4682}{{\tt
  0907.4682}}].

\bibitem{Ellwanger:2005dv}
U.~Ellwanger and C.~Hugonie, \emph{{NMHDECAY 2.0: An Updated program for
  sparticle masses, Higgs masses, couplings and decay widths in the NMSSM}},
  \href{http://dx.doi.org/10.1016/j.cpc.2006.04.004}{\emph{Comput. Phys.
  Commun.} {\bf 175} (2006) 290--303},
  [\href{https://arxiv.org/abs/hep-ph/0508022}{{\tt hep-ph/0508022}}].

\bibitem{Ellwanger:2004xm}
U.~Ellwanger, J.~F. Gunion and C.~Hugonie, \emph{{NMHDECAY: A Fortran code for
  the Higgs masses, couplings and decay widths in the NMSSM}},
  \href{http://dx.doi.org/10.1088/1126-6708/2005/02/066}{\emph{JHEP} {\bf 02}
  (2005) 066}, [\href{https://arxiv.org/abs/hep-ph/0406215}{{\tt
  hep-ph/0406215}}].

\bibitem{Huang:2014xua}
P.~Huang and C.~E.~M. Wagner, \emph{{Blind Spots for neutralino Dark Matter in
  the MSSM with an intermediate $m_A$}},
  \href{http://dx.doi.org/10.1103/PhysRevD.90.015018}{\emph{Phys. Rev.} {\bf
  D90} (2014) 015018}, [\href{https://arxiv.org/abs/1404.0392}{{\tt
  1404.0392}}].

\bibitem{Baum:2017gbj}
S.~Baum, K.~Freese, N.~R. Shah and B.~Shakya, \emph{{NMSSM Higgs boson search
  strategies at the LHC and the mono-Higgs signature in particular}},
  \href{http://dx.doi.org/10.1103/PhysRevD.95.115036}{\emph{Phys. Rev.} {\bf
  D95} (2017) 115036}, [\href{https://arxiv.org/abs/1703.07800}{{\tt
  1703.07800}}].

\bibitem{Baum:2018zhf}
S.~Baum and N.~R. Shah, \emph{{Two Higgs Doublets and a Complex Singlet:
  Disentangling the Decay Topologies and Associated Phenomenology}},
  \href{https://arxiv.org/abs/1808.02667}{{\tt 1808.02667}}.

\bibitem{Baum:2019uzg}
S.~Baum, N.~R. Shah and K.~Freese, \emph{{The NMSSM is within Reach of the LHC:
  Mass Correlations \& Decay Signatures}},
  \href{http://dx.doi.org/10.1007/JHEP04(2019)011}{\emph{JHEP} {\bf 04} (2019)
  011}, [\href{https://arxiv.org/abs/1901.02332}{{\tt 1901.02332}}].

\bibitem{Baum:2019pqc}
S.~Baum and N.~R. Shah, \emph{{Benchmark Suggestions for Resonant Double Higgs
  Production at the LHC for Extended Higgs Sectors}},
  \href{https://arxiv.org/abs/1904.10810}{{\tt 1904.10810}}.

\bibitem{Gori:2018pmk}
S.~Gori, Z.~Liu and B.~Shakya, \emph{{Heavy Higgs as a Portal to the
  Supersymmetric Electroweak Sector}},
  \href{http://dx.doi.org/10.1007/JHEP04(2019)049}{\emph{JHEP} {\bf 04} (2019)
  049}, [\href{https://arxiv.org/abs/1811.11918}{{\tt 1811.11918}}].

\bibitem{Adhikary:2018ise}
A.~Adhikary, S.~Banerjee, R.~Kumar~Barman and B.~Bhattacherjee, \emph{{Resonant
  heavy Higgs searches at the HL-LHC}},
  \href{https://arxiv.org/abs/1812.05640}{{\tt 1812.05640}}.

\bibitem{Basler:2018dac}
P.~Basler, S.~Dawson, C.~Englert and M.~Mühlleitner, \emph{{Showcasing HH
  production: Benchmarks for the LHC and HL-LHC}},
  \href{http://dx.doi.org/10.1103/PhysRevD.99.055048}{\emph{Phys. Rev.} {\bf
  D99} (2019) 055048}, [\href{https://arxiv.org/abs/1812.03542}{{\tt
  1812.03542}}].

\end{thebibliography}\endgroup
\bibliographystyle{JHEP}
 
\end{document}